\documentclass[accepted]{smartcom_ww}
\usepackage{latexsym}

\usepackage{ifpdf}
\usepackage{cite}
\usepackage{array}
\usepackage{stfloats}
\usepackage[dvips]{graphics}
     \usepackage{color}
     \usepackage{graphicx}
    \usepackage[cmex10]{amsmath}
    \usepackage{float}
\usepackage{cite}
\usepackage{multirow}
\usepackage{hhline}
    \usepackage{adjustbox}
    \usepackage{booktabs}
    \usepackage{xparse}
\usepackage{flushend}
\etitle{\huge{Dynamic Resource Allocation and Activity Management for Energy Efficiency and Fairness in Heterogeneous Networks}}
\authorlist{%
 \authorentry[amir.behrouzifar@rutgers.edu]{}{Amir Behrouzi-Far}{Rutgers}%
 \authorentry[ezhan@ee.bilkent.edu.tr]{}{Ezhan Kara\c{s}an}{Bilkent}%
}
\affiliate[Rutgers]{}{Department of Electrical and Computer Engineering, Rutgers University, Piscataway, NJ 08854, USA}
\affiliate[Bilkent]{}{Department of Electrical and Electronics Engineering, Bilkent University, Ankara 06800, Turkey}

\begin{document}
\begin{eabstract}
Higher energy consumption of Heterogeneous Networks (HetNet), compared to Macro Only Networks (MONET), raises a great concern about the energy efficiency of HetNets. In this work we study a dynamic activation strategy, which changes the state of small cells between Active and Idle according to the dynamically changing user traffic, in order to increase the energy efficiency of HetNets. Moreover, we incorporate dynamic inter-tier bandwidth allocation to our model. The proposed Dynamic Bandwidth Allocation and Dynamic Activation (DBADA) strategy is applied in cell-edge deployment of small cells, where HotSpot regions are located far from the master base station. Our objective is to maximize the sum utility of the network with minimum energy consumption. To ensure proportional fairness among users, we used logarithmic utility function. To evaluate the performance of the proposed strategy, the median, 10-percentile and the sum of users' data rates and the network energy consumption are evaluated by simulation. Our simulation results shows that the DBADA strategy improves the energy consumed per unit of users' data rate by up to $25\%$. It also achieves lower energy consumption by at least $25\%$, compared to always active scenario for small cells.
\end{eabstract}
\begin{ekeyword}
Heterogeneous Network, Small Cell, Spectrum Sharing, Cell Edge, Energy Efficiency, Fairness.
\end{ekeyword}
\maketitle

\section{Introduction}
The growth in the number of mobile users and the increase in their required data rates motivate researchers
 to think about more advanced wireless network architectures \cite{zhang20155g,wang2014cellular}. The new architectures should be able to cope with the conventional network planning, where a single Base Station (BS) is responsible for coordinating all the communications that take place within a large cell. Among all, a heterogeneous deployment of small BSs (SBSs) which is underlaid by a conventional single-BS network is proved to be promising for improving both the network coverage and the users quality of experience. \cite{ghosh2012heterogeneous,chandrasekhar2008femtocell}.

The direct result of deploying a large number of BSs inside a conventional single-BS cell is an inevitable growth in
the network energy consumption. Therefore, since the emergence of HetNets, there has been a great concern
 about their energy efficient deployments \cite{soh2013energy,baker2012lte,nakamura2013trends}. However, according
 to \cite{feng2016boost}, there are great fluctuations in the number of active users in a cellular network. Those fluctuations, which are
 experienced over time and space, enable us to think about more efficient deployments of small cells. For instance, putting small cells to Idle state in off-peak hours is proposed in the literature \cite{gougeon2015energy}. In this strategy, a small cell is pushed to Idle state if its serving traffic drops below a certain threshold and would switch
 back to Active state if its traffic passes a (not necessarily the same) threshold. The two thresholds might not be the same to prevent unnecessary instantaneous
 switches. When a small cell goes to Idle state, its users would be offloaded to either Macro BS (MBS) or by other SBSs \cite{wang2015context}.

There is a diverse literature about energy efficient deployment of HetNets. Authors in \cite{cai2016green} propose a dynamic ON/OFF strategy for both uniform and non-uniform user distributions. In this work, the achievable degree of energy efficiency under a certain outage probability is studied. However, the effect of energy saving on the potential improvements of HetNets over Macro Only Networks (MONETs) has not been studied. In \cite{liu2015flow}, the traffic delay and energy efficiency trade-off in HetNets is studied. A dynamic ON/OFF strategy over a dynamically changing traffic in an HetNet is studied in \cite{tran2015dynamic}, where the authors try to maximize energy efficiency of the network. Although their sum rate analysis sheds some light on the possible energy efficiency improvement of HetNets, they do not consider fairness among users. A joint partial spectrum sharing and ON/OFF strategy for a two tier HetNet is introduced in \cite{chai2015joint}. In this work, the coverage probabilities with different reuse factors are examined. An activity management algorithm, based on state switching of small cells, is proposed in \cite{aykin2019activity}.

In this paper, we study the problem of energy efficient deployment of HetNets. we study a dynamic activation and dynamic spectrum allocation strategy, with the objective of maximizing the sum utility and energy efficiency of the network. By using a logarithmic utility function, we aim at providing proportional fairness among the active users in the network. In order to consider the price of energy, we introduce a new parameter, which essentially adjusts the weight of energy efficiency in the optimization problem. We evaluate the improvement of DBADA strategy by studying the sum, median and 10-percentile of users' data rate. We show, by simulation, that by employing DBADA with Proportional Fair Scheduling (PFS), both energy efficiency and users' experience could be considerably improved.

The rest of this paper is organized as follows. In Section II we present our system model. The optimization problem and discussions about the solutions are provided in Section III. The simulations result and conclusions are presented in Section IV and Section V, respectively.

\section{system model}
In this work, a two tier HetNet is considered. As depicted in Fig. 1, small cells are deployed far from MBS to improve cell edge experience. User associations are done based on the received SNR, such that a user would be assigned to macro tier if its received SNR from MBS is higher than all the SBSs. Otherwise it will be assigned to a small cell which provides the highest SNR. We dynamically divide spectrum among different tiers/cells, in a way that it gives the highest sum utility in the network. The utility of a user is logarithmically proportional to its data rate. This choice of utility function ensures fairness among users, by preventing the exclusive assignment of spectrum to the closely located users.

In our dynamic Active/Idle switching strategy, we assume that idle small cells consume a fixed amount of energy, for staying coordinated with the rest of network, which is much smaller than the energy they consume when they are active. Our user distribution model is according to the HotSpot model, where on top of the uniform distribution of users, some HotSpot regions are also existed in the area. The HotSpot regions are considered at the cell edge of the macro cell, Fig 1. In order to capture the price of energy, we introduce a new parameter, $0\leq\beta\leq 1$, which essentially adjusts the importance of the energy saving in the joint problem.

\begin{figure}[!t]
\centering
\includegraphics[height=9cm, width=7cm]{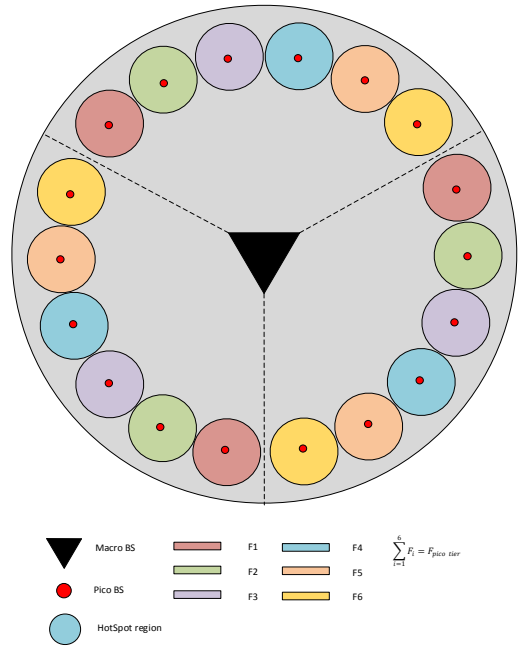}
\caption{Network architecture}

\end{figure}

\section{problem statement and solution}
The problems of utility maximization and dynamic activation have been separately studied in the literature. In this paper, we study the joint problem, in order to understand the trade-off between energy efficiency and users' experience. Thus our optimmization problem is written as,
\begin{equation}
\begin{split}
max   \qquad &\sum_{n=1}^{N} U(r_n) - \beta \sum_{k=1}^{K} E_k\\
 &s.t \quad \sum_{n=1}^{N} W_n = W_T,
\end{split}
\end{equation}
where $r_n$ is the data rate and $U(r_n)$ is the utility experienced by user $n$ and $N$ is the total number of users in the system. $E_k$ is the energy consumed by $k$th BS and $K$ is the total number of BSs. Here $\beta$ corresponds to the relative price of energy in the studied environment. $W_n$ and $W_T$ are the spectrum share of $n$th user and the total available spectrum in the system, respectively. The data rate of user $n$ is given by,
\begin{equation}
\begin{split}
r_n = W_n log (1+\frac{P_S}{N_0W_n}).
\end{split}
\end{equation}

There are a broad range of utility functions defined in the literature. Authors in \cite{irwin2004balancing} use a linear utility function which is suitable for FTP file transmission. Another utility function, which is known to guarantee proportional fairness in the system \cite{kelly1998rate}, is the logarithmic function. The basic idea behind using logarithmic utility functions is that, due to their concavity, they allocate higher spectrum to the penalized users. This simple but important feature of logarithmic functions makes them a good feat for improving fairness in communication systems \cite{kelly1998rate}, where the goal is to provide an acceptable quality of experience for most of the users. Accordingly, the problem of an energy-efficient proportional fair 2-tier HetNet can be written as
\begin{equation}
\begin{split}
max   \qquad &\sum_{n=1}^{N} ln(r_n) - \beta \sum_{k=1}^{K} E_k\\
 &s.t \quad \sum_{n=1}^{N} W_n = W_T,
\end{split}
\end{equation}
where $ln$ stands for the natural algorithm.
Due to the concavity of $ln(.)$ function, the objective function (3) is a concave function of $W_n$ \cite{behrouzi2016dynamic}. Therefore, the problem has only one global solution. Considering $K$ BSs inside the network, there are $2^K$ distinct concave problems that need to be solved. After solving the set of all possible optimization problems there are $2^K$ solutions, each of which associates with a specific BSs state vector. The BSs state vector is a binary vector the length of which is equal to the number of BSs. In this binary vector, 1s stand for Active state and 0s stand for Idle state. Finally, the optimum solution of the problem is the one associated with the maximum objective between all $2^K$ objectives.

The parameter $\beta$, is responsible for unit compatibility of the two terms in the objective function as well as the relative price of the energy of the studied environment. Essentially, the larger $\beta$ stands for the greater price of the energy and the smaller $\beta$ stands for lower energy price. Therefore, it is expected that DBADA strategy pushes more small cells to Idle state as the value of $\beta$ gets larger.

\section{numerical results}
With the same model as in Fig. 1, 6 HotSpot regions per sector, are considered at the cell edge of macro cell. A macro cell consists of 6 sectors. Users' location is uniformly generated,
\begin{itemize}
  \item One over the entire Macro cell, with average number of users of $\overline{N}_m,$
  \item And six others over HotSpot regions, with the total average number of users of $\overline{N}_h.$
\end{itemize}

In addition, the fluctuations of the number of active users, are modeled with zero-mean uniform random variables, $n_m$ and $n_h$. Therefore, the instantaneous number of users in macro cell and HotSpot are given by,
\begin{equation}
\begin{split}
    & N_m = \overline{N}_m+n_m \\
     & N_h = \overline{N}_h+n_h.
\end{split}
\end{equation}
We considered dynamic cahnges in the average number of active users over time. Specifically, we considered,
\begin{itemize}
  \item $\overline{N}_m=[197, 170, 140, 110, 80, 50, 20, 5, 5],$
  \item $\overline{N}_h=[1, 10, 20, 30, 40, 50, 60, 65, 65],$
\end{itemize}
where each element of the voctors are the average number of users. The other simulation parameters are summarized in Table I. In our numerical analysis the following three scenarios are implemented,
\begin{itemize}
  \item \emph{Macro Only (MO)}, in which all pico cells are in Idle state and all the users are served by macro cell.
  \item \emph{Picos Active with $\alpha$ percent BW $(PA\alpha\%)$}, in which all the pico cells are Active and are collectively using $\alpha$ percent of resources. Macro cell in these strategies uses $(100-\alpha)$ percent of resources.
  \item \emph{DBADA}, in which the optimum state of each pico cell (Active or Idle) is dynamically determined. Further, inter-tier and inter-cell resource allocations are dynamically adjusted according to traffic variations.
\end{itemize}
We consider 2 spectrum scheduling policies,
\begin{itemize}
    \item \emph{Proportional Fair Scheduling (PFS)}, in which the available spectrum is allocated among tiers/users such that the network sum utility maximized.
    \item \emph{Equal Allocation (EA)}, in which users get equal share of spectrum.
\end{itemize}
\begin{table}[!t]
\centering
\renewcommand{\arraystretch}{1.3}
\caption{Simulation parameters}
\begin{tabular}{|c|c|}
  \hline
   Parameter  & Value \\
  \hline
  \multirow{2}{*}{Consumed power} & Macro (per sector): 390W \\
    \hhline{~-}
    &Pico: 9W \\
    \hline
  \multirow{2}{*}{Air power} & Macro: 46dBm \\
    \hhline{~-}
    &Pico: 30dBm \\
    \hline
  \multirow{2}{*}{Path-loss model} & Macro: $128.1 +37.6log(R)$ \\
    \hhline{~-}
    &Pico: $140.7 +36.7log(R)$ \\
    \hline
  \multirow{2}{*}{Antenna gain} & Macro: 14dBi \\
    \hhline{~-}
    &Pico: 5dBi \\
    \hline
  \multirow{2}{*}{Antenna model} & Macro: sectorized with 3 sector \\
    \hhline{~-}
    &Pico: omni-directional \\
    \hline
  System bandwidth & 100MHz \\
  \hline
  Macro cell inter-site distance & 1000m \\
  \hline
\end{tabular}
\end{table}

The energy consumption of Macro only and Picos Active scenarios together with DBADA strategy are depicted in Fig. 2. During the off-peak hours of HotSpot traffic, when the users are almost uniformly distributed in the network, DBADA strategy pushes pico cells to Idle state. Specially, by increasing the value of $\beta$, which corresponds to the relative price of energy in the environment, DBADA strategy may push some pico cells to Idle state even in the HotSpot peak hours. This happens since, as discussed earlier, our model captures the fluctuation in the number of users in addition to the dynamic variation in the average serving traffic by each tier. Accordingly, DBADA may push a pico cell to Idle state if the cell experiences negative fluctuations which reduce its serving traffic. Moreover, by increasing $\beta$ it is expected that DBADA cares more about energy than the sum utility by pushing more pico cells to Idle state, which is shown by Fig. 2. On the other hand, as $\beta$ decreases the DBADA strategy gets closer to Picos Active strategy and tends to push even the lightly loaded pico cells to Active state.

\begin{figure}[htb]
\centering
\includegraphics[width=\columnwidth]{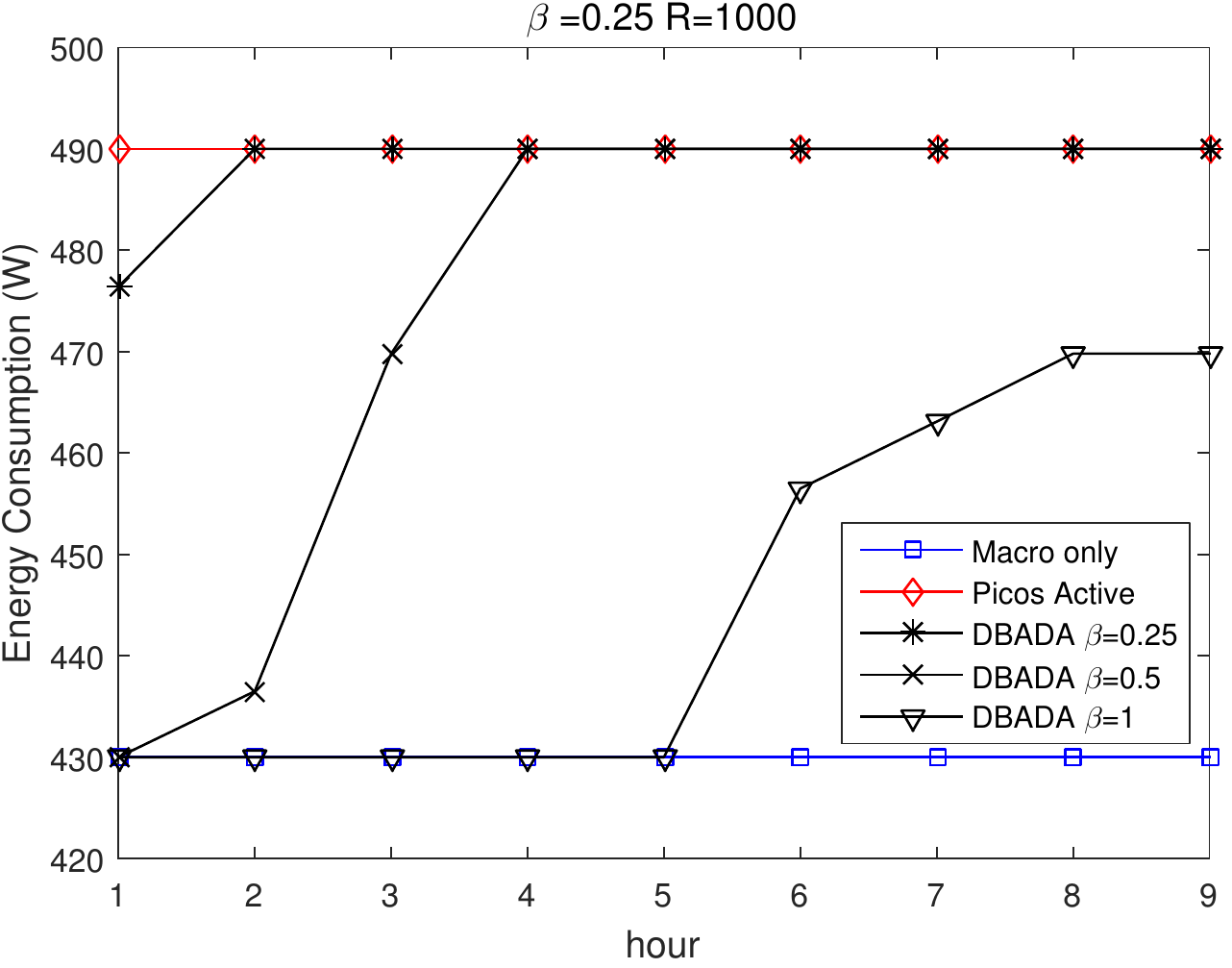}
\caption{Network energy consumption}

\end{figure}

\begin{figure}[!t]
\centering
\includegraphics[width=\columnwidth]{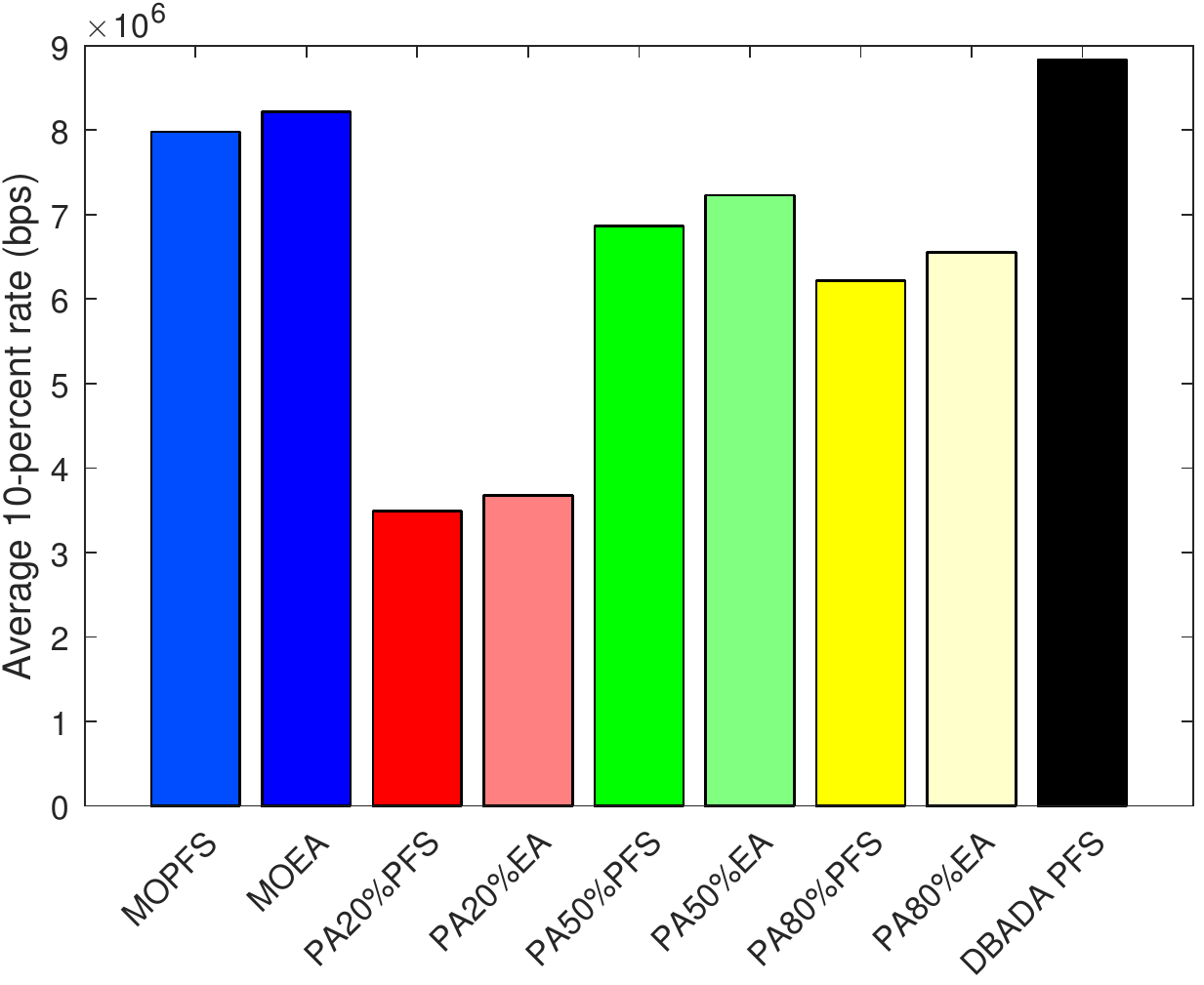}
\caption{Average of 10-percent rate, $\beta=0.5$.}

\end{figure}

\begin{figure}[!t]
\centering
\includegraphics[width=\columnwidth]{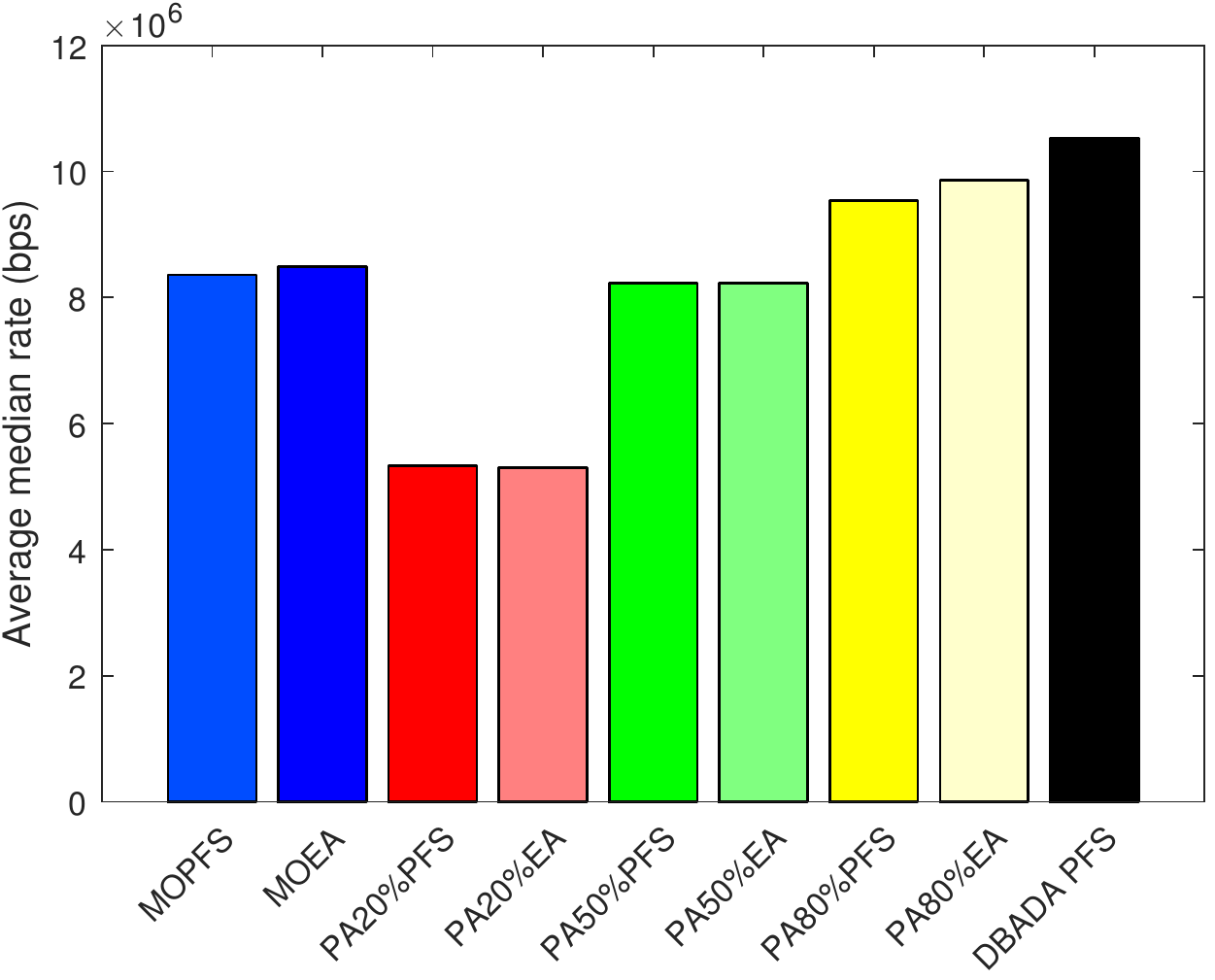}
\caption{Average of median rate, $\beta=0.5$.}

\end{figure}

\begin{figure}[!t]
\centering
\includegraphics[width=\columnwidth]{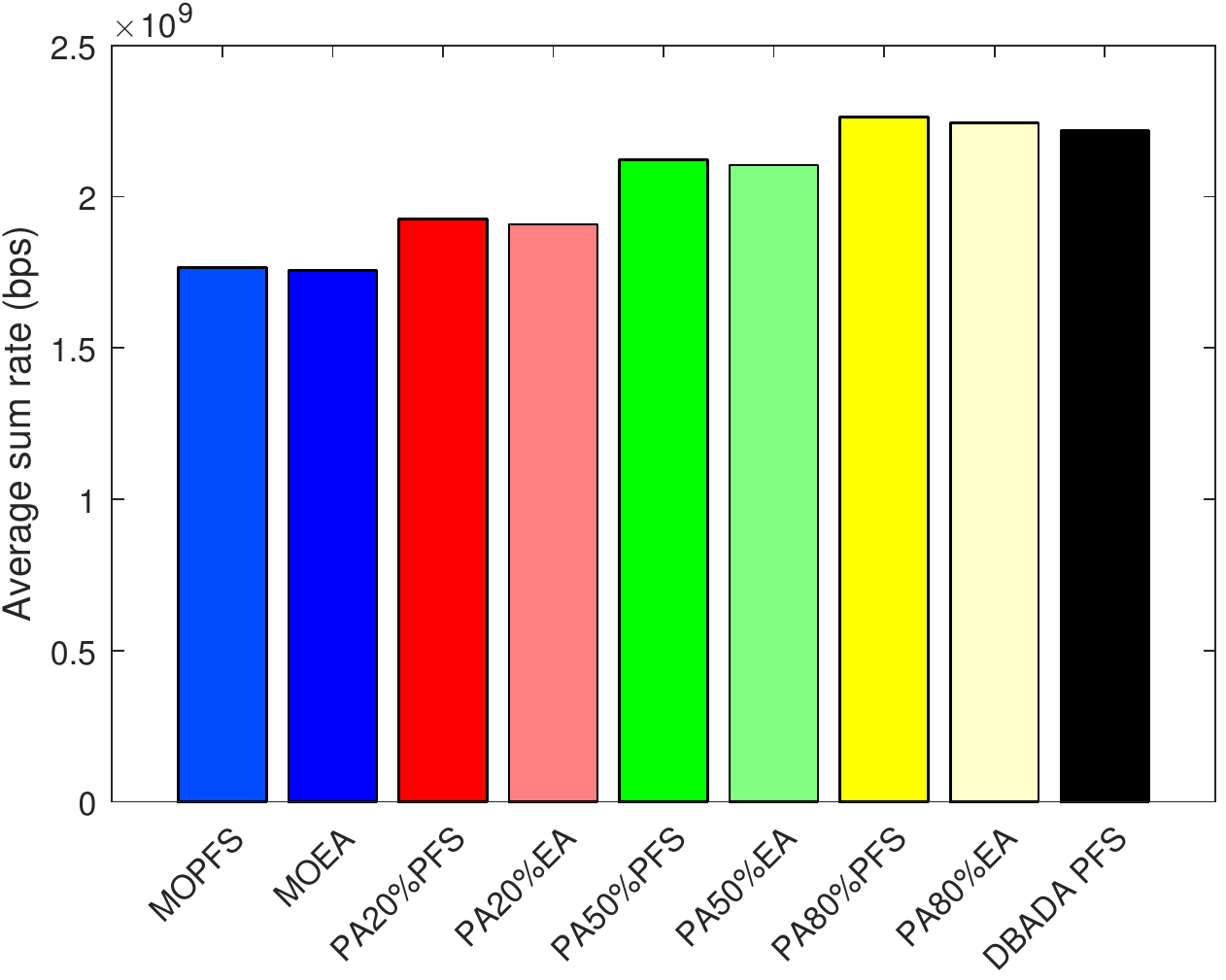}
\caption{Average of sum rate, $\beta=0.5$.}

\end{figure}

In Fig. 3, the average of 10-percent rates over nine hours of dynamic traffic is plotted. As it is shown, the DBADA strategy can improve the 10-percentile rate by at least $10\%$, compared to other strategies. The important observation in this figure is the MO scenario outperforms all the PA$\alpha\%$ scenarios with fixed inter-tier spectrum allocation. This result shows the importance of dynamic spectrum allocation when traffic changes dynamically. The proposed DBADA strategy improves the median rate at least by $10\%$, as shown in Fig. 4. Unlike the 10-percentile rate, MO is not a good strategy for improving median rate. PA$80\%$, on the other hand, performs very close to DBADA strategy. This behavior is in line with intuition, the small cell users are closer to the base station and allocating them higher spectrum will increase the median user experience. With sum rate as evaluation scheme, on the other hand, PA$80\%$ is the best strategy, since it keeps the small cells always active and allocates them high portion of spectrum. This will result the small cell user, which are also in advantage of being close to their BS, to gain high data rates and maximize the sum rate compared to other strategies. However, keeping small cell always in active state is not energy efficient and reasonable trade-off between energy consumption and user experience is necessary.

In order to incorporate energy consumption into data rate metrics, we introduce the average energy per sum, median and 10-percentile rates. These metrics capture the amount of energy spent per unit of the data rate metrics. In Table II, the achievable improvement of DBADA strategy over other scenarios are summarized. While achieving the least energy per network sum rate, DBADA strategy, compared to PA$\alpha\%$ scenarios, consumes at least $26\%$ and $21\%$ less energy per median and 10-percentile rates, respectively. Furthermore, by achieving above $10\%$ improvement in energy per median rate metric over MO scenario, DBADA scenario guarantees better energy efficiency for median users. Nevertheless, although DBADA strategy achieves higher 10-percentile rate than MO scenario, in terms of energy per 10-percentile rate it performs the same as MOPFS, and $4\%$ less than MOEA.

\NewDocumentCommand{\rot}{O{290} O{5em} m}{\makebox[#2][l]{\rotatebox{#1}{#3}}}%
\begin{table}[!t]
\centering
\renewcommand{\arraystretch}{1.3}
\caption{Improvement of DBADA over other scenarios, $\beta=0.5$.}
\begin{tabular}{|c|c|c|c|}
  \hline
  &  \rot{Avg. energy/sum rate} & \rot{Avg. energy/median rate} & \rot{Avg. energy/10\% rate} \\
  \hline
  MOPFS & 12\% & 12\% & 0\% \\
  \hline
  MOEA & 13\% & 10\% & -4\% \\
  \hline
  PA20\%PFS & 16\% & 63\% & 65\% \\
  \hline
  PA20\%EA & 53\% & 62\% & 63\% \\
  \hline
  PA50\%PFS & 7\% & 27\% & 26\% \\
  \hline
  PA50\%EA & 8\% & 26\% & 21\% \\
  \hline
  PA80\%PFS & 1\% & 29\% & 45\% \\
  \hline
  PA80\%EA & 2\% & 28\% & 42\% \\
  \hline
\end{tabular}
\end{table}

\section{Conclusion}
A dynamic resource allocation and dynamic activation strategy for HetNets was proposed. The optimization problem, with the objective of maximizing network utility and energy efficiency was formulated. In the studied scenario small cells were employed far from MBS to improve cell edge experience. Through simulations, we observed that DBADA strategy achieves highest average median and average 10-percent rates while maintaining close-to-maximum average sum rate. In addition, with the minimum energy per network sum rate and median rate, DBADA was shown to achieve at least $21\%$ improvement in energy per 10-percentile rate over small cells-always-active scenarios.
\bibliography{whole}
\bibliographystyle{ieeetr}
\end{document}